\documentclass[12pt,preprint]{aastex}

\lefthead{Chou et al.}

\righthead{Abundances of the TriAnd Overdensity}

\begin{document}
\title{First Chemical Analysis of Stars in The Triangulum-Andromeda Star Cloud}

\author{Mei-Yin Chou\altaffilmark{1,2},
Steven R. Majewski\altaffilmark{1}, Katia
Cunha\altaffilmark{3,4,7},
Verne V. Smith\altaffilmark{3}, \\
Richard J. Patterson\altaffilmark{1}, and David
Mart{\'i}nez-Delgado\altaffilmark{5,6}
}

\altaffiltext{1}{Dept. of Astronomy, University of Virginia,
Charlottesville, VA 22904-4325 (mc6ss, srm4n, rjp0i@virginia.edu)}

\altaffiltext{2}{Institute of Astronomy and Astrophysics, Academia
Sinica, Taipei 10617, Taiwan (cmy@asiaa.sinica.edu.tw)}

\altaffiltext{3}{National Optical Astronomy Observatories, PO Box
26732, Tucson, AZ 85726 (cunha, vsmith@noao.edu)}

\altaffiltext{4}{Steward Observatory, University of Arizona,
Tucson, AZ 85721, USA}

\altaffiltext{5}{Instituto de Astrofisica de Canarias, La Laguna,
Spain (ddelgado@iac.es)}

\altaffiltext{6}{Max-Planck-Institut f\"{u}r Astronomie,
K\"{o}nigstuhl 17 , Heidelberg, Germany }

\altaffiltext{7}{On leave from Observatorio Nacional, Rio de
Janeiro, Brazil}

\begin{abstract}

We undertake the first high resolution spectroscopic study of the
Triangulum-Andromeda (TriAnd) star cloud --- an extended,
mid-latitude Milky Way halo substructure about 20 kpc away in the
second Galactic quadrant --- through six M giant star candidates
selected to be both spatially and dynamically associated with this
system. The abundance patterns of [Ti/Fe], [Y/Fe] and [La/Fe] as a
function of [Fe/H] for these stars support TriAnd as having an
origin in a dwarf galaxy with a chemical enrichment history
somewhat similar to that of the Sagittarius dwarf spheroidal
(dSph) galaxy. We also investigate the previously proposed
hypothesis that TriAnd is an outlying, dynamically older piece of
the Monoceros Stream (also known as the Galactic Anticenter
Stellar Structure, ``GASS'') under the assumption that both
features come from the tidal disruption of the same accreted Milky
Way satellite and find that net differences in the above abundance
patterns between the TriAnd and GASS stars studied suggest that
these two systems are independent and unrelated.

\end{abstract}


\section{Introduction}

The Triangulum-Andromeda (TriAnd) star cloud was identified as a
low-latitude overdensity of M giant stars in the Two Micron
All-Sky Survey (2MASS) by Rocha-Pinto et al. (2004, hereafter
RP04) and as a strong sequence of main sequence turn-off (MSTO)
stars by Majewski et al. (2004, hereafter M04). RP04 and M04 found
the structure to extend across at least
$50^{\circ}\times20^{\circ}$ of the halo at a heliocentric
distance of 15-30 kpc. The mean heliocentric radial velocity of
the structure is $\sim-119\pm9.6$ km s$^{-1}$, and its mean [Fe/H]
is $\sim-1.2$ dex based on calcium infrared triplet measurements
by RP04. Though a subsequent MegaCam survey of MSTO stars in this
sky region (Martin, Ibata, \& Irwin 2007) confirmed a $\sim20$ kpc
distance for this stellar structure, to date little additional
attention has been given to TriAnd, including determining its
properties at any level of detail useful for ascertaining the
origin of the structure.

In contrast, significantly more attention has been given to the
Galactic Anticenter Stellar Structure (GASS), also called the
``Monoceros Stream'' or ``Monoceros Ring'', which was discovered
only two years before TriAnd as several overdensities of presumed
MSTO stars (Newberg et al. 2002) in the same general region of the
Galactic anticenter as TriAnd\footnote{Actually GASS/Monoceros was
originally seen in Monoceros, but soon mapped much more widely
including across the constellations Triangulum and Andromeda; see
references in following text.} but at a closer mean $\sim11$ kpc
heliocentric distance. Subsequent study of this ring-like
structure with the Isaac Newton Telescope Wide Field Camera (Ibata
et al. 2003; Conn et al. 2005), Sloan Digital Sky Survey
spectroscopy and photometry (Yanny et al. 2003, hereafter Y03),
and as traced by 2MASS M giants (Crane et al. 2003, hereafter C03;
Majewski et al. 2003, hereafter M03; Rocha-Pinto et al. 2003,
hereafter RP03) has shown that the low latitude GASS ring spans at
least the second and third Galactic quadrants and has a wide
metallicity spread, from [Fe/H]= $-1.6\pm 0.3$ (Y03) to $-0.4\pm
0.3$ (C03). Nevertheless, the origin of GASS remains
controversial. Principal scenarios include those where GASS is a
piece/warp of the Galactic disk (Momany et al. 2006; Kazantzidis
et al. 2008; Younger et al. 2008) versus those where it represents
tidal debris from the disruption of a Milky Way (MW) satellite
galaxy (Ibata at al. 2003; Y03; C03; RP03; Helmi et al. 2003;
Frinchaboy et al. 2004; Martin et al. 2004; Pe{\~{n}}arrubia et
al. 2005; Conn et al. 2005; Rocha-Pinto et al. 2006). A comparison
of chemical patterns between the GASS and other dwarf spheroidal
(dSph) galaxies has recently been made in Chou et al. (2010b,
hereafter C10b). Their GASS sample reveals similar abundance
patterns for Ti, Y and La as in the Sagittarius dwarf+stream
system and other satellite galaxies, which suggests that stars in
GASS likely formed in a dwarf galaxy-like environment. However,
this chemical observation does not in itself conclusively resolve
the debate over the {\it dynamical} origin of GASS because
theoretical models show that galactic disks grow outward by
accretion of dwarf galaxies (e.g., Abadi et al. 2003; Yong,
Carney, \& Teixera de Almeida 2005), and whether GASS represents
an intact tidal stream versus some other Galactic structure
generated from a previously hierarchically-formed, outer disk is
not clarified by this chemistry.

The lack of a definitive origin model for GASS notwithstanding,
soon after the TriAnd discovery it was proposed by Pe\~{n}arrubia
et al. (2005) that this structure may simply be an outlying piece
of GASS/Mon, which these authors hypothesized to be a structure
formed from the dissolution of a MW satellite galaxy. Both
structures are at similar Galactic latitudes, and Pe\~{n}arrubia
et al. were able to find an $N$-body simulation that could
incorporate TriAnd as a dynamically older, earlier-stripped piece
from the putative Mon progenitor galaxy, under the assumption that
the position of this progenitor is in Canis Major (the latter
aspect motivated by the claim of a Canis Major overdensity
interpreted as the core of a disrupting dSph by Martin et al. 2004
--- itself a greatly debated issue; see Rocha-Pinto et al. 2006,
Mateu et al. 2009, and references therein). If TriAnd and GASS/Mon
are parts of the same dwarf galaxy system, then the stars in these
two structures share a common chemical enrichment history and
should in principle show similar enrichment patterns -- e.g.,
those seen by C10b for GASS/Mon. In effect, one could apply
``chemical fingerprinting'' to determine whether the stars in
TriAnd belong to the GASS/Mon system.

This technique was recently applied by Chou et al. (2010,
hereafter C10a) to show that a group of M giant stars in the North
Galactic Hemisphere (the ``NGC moving group'') are likely older
pieces of the Sagittarius (Sgr) tidal stream. A comparison to the
Sgr system is quite apropos in the study of Mon and TriAnd, since
Sgr exhibits many similar properties to both, including a
relatively high metallicity with the presence of M giant stars in
all three cases (M03; RP03; RP04) that suggest the progenitor
systems of all three substructures may have been somewhat similar.

Therefore, following the above precedents from our previous
chemical studies of halo substructure, in this paper we make the
first high resolution spectroscopic assessment of the chemical
abundance patterns in the TriAnd system, and use these abundance
measurements to investigate the hypothesized connection between
the TriAnd and GASS/Mon systems.

\section{Observations and Analysis}

Our analysis of TriAnd stars is identical to that used for Sgr and
GASS M giants described in C07, C10a and C10b. We use spectra from
echelle spectrographs mounted on the Apache Point Observatory
3.5-m\footnote{The Apache Point Observatory 3.5-meter telescope is
owned and operated by the Astrophysical Research Consortium.} and
Kitt Peak National Observatory 4-m telescopes, with resolutions $R
= 32,000$ and 35,000, respectively. We measure equivalent widths
(EWs) of eleven \ion{Fe}{+1} lines, two \ion{Ti}{+1} lines, and
one \ion{Y}{+2} line, and use spectral synthesis for one
\ion{La}{+2} line in the spectral region 7440-7590 \AA\ previously
studied by Smith \& Lambert (1985; 1986; 1990). We use the LTE
code MOOG (Sneden 1973) to derive abundances of these elements
following the same procedures as in C07, C10a and C10b, and adopt
the same model atmospheres, generated by linear interpolation from
the grids of ATLAS9 models in Castelli \& Kurucz
(2003).\footnote{From http://kurucz.harvard.edu/grids.html.} Our
analysis used their ODFNEW models without convective overshooting.

We observed six TriAnd M giants selected to have the photometric
parallax and radial velocity of that system from the sample
described in RP04. Observed GASS stars range from $7.8 < K_s <
10.1$, while the TriAnd stars span a narrower, $10.5 < K_s < 11.2$
range. Nevertheless, both samples have the same $J-K_s = 1.0$ blue
limit and about the same intrinsic RGB luminosity, as evidenced by
a comparison to the fitted isochrones (see \S3). The primary
difference is that the GASS sample includes more very red,
presumably asymptotic giant branch (AGB) stars so that the red
limits are $(J-K_s)_o < 1.10$ for TriAnd and $ (J-K_s)_o < 1.25$
for GASS; this yields mean $(J-K_s)_o$ colors of 1.05 and 1.09 dex
for the full TriAnd and GASS samples, respectively. The apparent
$K_s$ magnitudes and dereddened $(J-K_s)_o$ colors (Schlegel et
al. 1998) for our target TriAnd stars are listed in Table 1. The
same information for GASS stars is found in C10b.

The derived abundances are also summarized in Table 1. Table 1
columns give the derived effective temperature using the
Houdashelt et al. (2000) color-temperature relation applied to the
2MASS $(J-K_s)_o$ color, and the derived values of the surface
gravity ($\log{g}$), microturbulence ($\xi$), abundance $A$(X),
and abundance ratios [Fe/H] or [X/H] for each element X as well as
the standard deviation in the abundance determinations. The latter
represent the line-to-line scatter (for Fe, Ti and Y) for the EW
measures, or different continuum level adjustments (for La), as
discussed in C07 and C10a.

Uncertainties in abundance ratios are derived as in C10b. Briefly,
we account for both uncertainties in measured EWs as well as
propagated uncertainties from the adopted stellar parameters. The
latter mainly come from the color-calibrated effective
temperatures and imprecise knowledge of the proper isochrones to
use for each star. The net uncertainties are estimated to be 0.14,
0.20, 0.19, 0.16 for [Fe/H], [Ti/Fe], [Y/Fe] and [La/Fe],
respectively.

\section{Results}

Figure 1 shows color-magnitude diagrams (CMDs) of the TriAnd and
GASS M giant candidates.  Based on our derived [Fe/H] for each
star (indicated by the color scale) we roughly matched these
distributions with a set of Marigo et al. (2008) isochrones for 8
Gyr, [$\alpha$/Fe]=0, RGB/AGB stars. The choice of 8 Gyr matches
the approximate age of Sgr stars of [Fe/H] $\sim-0.7$ (Siegel et
al. 2007); analogy with Sgr is the only information available to
inform such a decision for the systems of interest here. The
corresponding isochrone distances are $\sim$12 kpc (GASS) and
$\sim$22 kpc (TriAnd); these are consistent with previous
estimates of the distances to these structures. However, due to
the age-metallicity degeneracy of RGB isochrones, other
combinations of age and metallicity also fit reasonably well ---
e.g., 5 Gyr old isochrones at 25 kpc and 14 kpc also fit the
TriAnd and GASS distributions, respectively. The wider CMD spread
of GASS stars having similar metallicities to each other (e.g.,
[Fe/H]$\sim$$-1$) suggests a potentially larger relative distance
spread ($\sim$11-22 kpc) for GASS stars compared to TriAnd if we
assume tight, monotonic age-metallicity relations in the two
systems.

Figure 2 compares derived distributions of [Ti/Fe], [Y/Fe], and
[La/Fe] as a function of [Fe/H] for our TriAnd, GASS/Mon (from
C10b) and Sgr stars (from C07, C10a). A strength of the following
comparison of TriAnd, GASS/Mon and Sgr stars is that not only are
our abundances for all three systems derived homogeneously using
the same methodology (same lines, same model atmospheres, same
derivation of stellar parameters, etc.), but the stars selected
for comparison span a very narrow range of stellar atmospheric
parameters ($\sim350$K in temperature and $\sim1.0$ dex in
$\log{g}$) so that the lines under study are forming under similar
atmospheric conditions. Linear fits to the MW distributions (see
C10a) are also shown in Figure 2. In contrast to the general
chemical similarity of GASS and Sgr (except for [La/Fe] at the
highest metallicities; see C10b), we note some larger differences
between TriAnd and these other two systems (Fig. 2). While the
mean trend of [Ti/Fe] versus [Fe/H] for the TriAnd stars is
somewhat similar to the mean trends for Sgr and GASS (though the
three of six sampled TriAnd stars at [Fe/H] $\sim -0.6$ fall well
below stars of similar metallicity in Sgr and GASS), the s-process
element patterns reveal more obvious differences. The mean level
of [Y/Fe] (middle panel) for TriAnd stars is about 0.2 dex higher
(or about 2.5 times the error in the mean for TriAnd) than that
for either Sgr or GASS stars at the same [Fe/H], and almost at the
solar enrichment level and near the mean for MW stars. And while
the [La/Fe] pattern of TriAnd stars closely resembles that of Sgr
stars (including one star at [Fe/H] = $-0.3$ with enhanced
[La/Fe], hinting at a possible upturn at high metallicity), it is
less similar to, and slightly enhanced compared to, the GASS
pattern, at least in the mean.

To put this comparison on a more quantitative footing, we apply
the statistical method of bootstrap sampling (e.g., Efron 1979;
Efron \& Tibshirani 1993) to judge the differences between TriAnd
and GASS. For a more conservative test, we combine the $6+21$ data
points of TriAnd+GASS stars for [Ti/Fe] and [Y/Fe], and $5+19$ for
[La/Fe] as if all stars are from one dSph system. We then randomly
pick six stars from the 27 TriAnd+GASS star sample, and compute
and compare the mean values for [Ti/Fe], [Y/Fe] and [La/Fe] for
the six selected and 21 non-selected stars (or, in the case of
[La/Fe], five from 24 stars). This process is repeated with 10,000
random selections of six (five) stars. The derived distributions
of the bootstrap samplings for Ti, Y and La are approximately
normally distributed, and we find that there are only $\sim4.5\%$
with mean [Ti/Fe] less than the mean value of TriAnd stars in the
10,000 bootstrap samples, and $\sim3.3\%$ with mean [Y/Fe] and
$\sim2.5\%$ with mean [La/Fe] more than those in TriAnd stars. If
the bootstrap sampling is repeated with the same procedure as
above but for only GASS stars, we find only $\sim0.95\%$ with mean
[Ti/Fe] less than TriAnd in the 10,000 bootstrap samples, and only
$\sim1.1\%$ with mean [Y/Fe] and 0 with mean [La/Fe] more than
TriAnd.

If we restrict the GASS stars to the same color range spanned by
the TriAnd stars ($1.00 \le [J-K_s]_o \le 1.07$) to minimize
metallicity selection biases, we find the mean [Fe/H] for eight
GASS stars to be $-0.52\pm0.09$ dex with a dispersion (standard
deviation) of 0.26 dex, compared with a mean [Fe/H] of
$-0.64\pm0.08$ dex with a dispersion of 0.19 dex for the six
TriAnd stars. Applying the bootstrap method for the more
conservative test we described above, we find that there are only
$\sim22.0\%$ with mean [Ti/Fe] less than the mean value of TriAnd
stars in the 10,000 bootstrap samples, and $\sim4.9\%$ with mean
[Y/Fe] and $\sim10.0\%$ with mean [La/Fe] more than those in
TriAnd stars for the combination of $6+8$ data points of
TriAnd+GASS stars for [Ti/Fe] and [Y/Fe], and $5+7$ for [La/Fe].
If we repeat the bootstrap sampling for the eight GASS stars only
and compare results with the mean values in TriAnd, we find only
$\sim4.4\%$ with mean [Ti/Fe] less than TriAnd in the 10,000
bootstrap samples, and 0 with mean [Y/Fe] and [La/Fe] more than
TriAnd, respectively.

Given the marked degree to which the mean values of {\it all
three} of the [Ti/Fe], [Y/Fe] and [La/Fe] ratios for TriAnd stars
are different from corresponding means for the various bootstrap
samples, we conclude that there might be significant chemical
differences between the two systems, sufficient to suggest that
TriAnd is not likely a part of the GASS system.

\section{Discussion}

As discussed in C10a, distinctive abundance patterns in stars
reflect the unique enrichment history of their parent system, and
therefore ``chemical fingerprinting'' can help identify tidally
stripped and captured stars in the Galactic field. The chemical
patterns of $\alpha$-elements (e.g., Ti) and s-process elements
(e.g., Y and La) may indicate differences in the (early) star
formation rate (SFR) between the progenitors of the TriAnd, Sgr,
and GASS systems, as well as the MW. For example,
$\alpha$-elements are mainly produced from Type II supernovae (SN
II) while iron is synthesized largely by Type Ia (SN Ia). So
[$\alpha$/Fe] is high in the early enrichment of a stellar system
until SN Ia occur after the first $\sim1$ Gyr. A shift to a lower
[$\alpha$/Fe] trend compared to the MW at some particular [Fe/H]
indicates a relatively slower initial SFR in that system, a
feature seen in many dSph galaxies as well as the LMC (e.g., Smith
et al.\ 2002, Shetrone et al.\ 2003; Tolstoy et al.\ 2003; Venn et
al.\ 2004; Geisler et al.\ 2005; Pomp{\'e}ia et al.\ 2008; C10a).
TriAnd follows the same $\alpha$ trends as dSph galaxies,
consistent with an origin in a dwarf galaxy-like environment and,
like other MW satellites, a low early SFR compared to the MW.

The s-process elements are primarily generated in low mass AGB
stars. In more metal-poor environments AGB stars produce heavier
s-process elements like La more efficiently than lighter species
such as Y (see review by Busso, Gallino, \& Wasserburg 1999). C10a
found underabundant trends in [Y/Fe] for all Sgr stars compared to
the MW, and an upturn in [La/Fe] for Sgr at [Fe/H]$>$$-0.5$.  The
latter trend also can be seen in McWilliam \& Smecker-Hane (2005),
and is an indication of a slow enrichment history, so that the
yields from low-metallicity AGB stars have enough time to
contaminate the interstellar medium and leave their signature on
the metal-rich populations (Venn et al.\ 2004; Pomp{\'e}ia et al.\
2008). The [Y/Fe] trends in TriAnd stars are slightly higher than
those in either Sgr and GASS and somewhat closer to MW trends,
whereas TriAnd shows a slightly lower [La/Fe] pattern compared to
Sgr. This suggests that AGB stars in the TriAnd progenitor
produced heavy s-process elements less efficiently than in Sgr ---
i.e., that low metallicity AGB progenitors in TriAnd contributed
relatively less heavy s-process enrichment than those in Sgr. This
may also indicate that the TriAnd progenitor was more metal-rich
than the Sgr progenitor. On the other hand, the one TriAnd star
with [Fe/H]$>$$-0.5$ shows slightly supersolar [La/Fe] (like the
Sgr [La/Fe] pattern, only less extreme), but above the GASS
[La/Fe] pattern at these metallicities. This may suggest that the
TriAnd system enriched more slowly than the progenitor(s) that
generated the GASS stars, but more similar in rate to the Sgr
dSph, to allow expression of the high La yields from those
relatively fewer low metallicity AGB stars that it did contain;
but this conclusion must be regarded as quite tentative, given the
errors on individual abundances and that this is based on only a
single high metallicity TriAnd star.

The $N$-body simulations from Pe\~{n}arrubia et al. (2005) suggest
that TriAnd may be an older piece of the Mon tidal stream.
According to the Sgr paradigm, older debris should be more
metal-poor than younger debris, but share the same chemical
pattern trends if from the same parent system. Indeed,
Pe\~{n}arrubia et al. associate older model stream wraps with
lower metallicity stars, but these data derive from lower
resolution spectroscopic measures of Ca II line indices for GASS
and TriAnd stars (C03, RP04). Though our study here has focused on
better precision, high resolution abundance determinations,
including [Fe/H] derived from eleven individual Fe line measures,
unfortunately the sample of stars is still too small to make
definitive conclusions about the relative metallicity of the two
systems.  On the other hand, the sample of TriAnd and GASS stars
in hand does offer tantalizing evidence that, at least at the
metallicities that were sampled, TriAnd does not reveal the same
chemical pattern trends as GASS and that therefore the TriAnd star
cloud is not part of an extended GASS system.\footnote{It is
perhaps worth noting that the inconsistency of velocities and
positions between stars in the TriAnd system (RP04) and those in
Sgr disruption models (e.g., Law, Johnston,
 \& Majewski 2005) or observed in the Sgr system (e.g., Majewski et
al. 2003, 2004) excludes a connection between the TriAnd and Sgr
systems, though both are rich in M giants.} That said, larger
samples of GASS and TriAnd stars having precisely determined
metallicities and chemical abundance patterns would certainly
provide a more conclusive chemical test of the possible connection
between these two systems, as well as more insights into the
origin and evolution of their respective progenitors.

\acknowledgements M.-Y.C., S.R.M. and R.J.P. acknowledge support
from NSF grants AST-0307851 and AST-0807945. This project was also
supported by the NASA {\it SIM Lite} key project {\it Taking
Measure of the Milky Way} under NASA/JPL contract 1228235. K.C.
and V.V.S. also appreciate support from NSF grant AST-0646790. D.
M.-D. acknowledges funding from the Spanish Ministry of Education
and Science (Ram{\'o}n y Cajal program contract and research
project AYA 2007-65090). Finally, we appreciate comments from two
anonymous referees that helped improve this paper.

\begin{deluxetable}{lr@{}rr@{}r@{}l@{}rrrrr@{}lrr@{}lr@{}lr}
\tabletypesize{\tiny} \tablecaption{Stellar Parameters and
Chemical Abundances for the Program Stars} \tablewidth{0pt}
\tablehead{ \colhead{Star No.} & \colhead{$\!K_{s,o}\,$} &
\colhead{$\,(J-K_s)_o\!$} & \colhead{$T_{\rm
eff}$\tablenotemark{a}} & \colhead{log{\it g}} & \colhead{} &
\colhead{$\xi$ } & \colhead{A(Fe)} &
\colhead{[Fe/H]\tablenotemark{b}} & \colhead{A(Ti)} &
\colhead{[Ti/Fe]\tablenotemark{b}}& \colhead{}& \colhead{A(Y)} &
\colhead{[Y/Fe]\tablenotemark{b}} & \colhead{} & \colhead{A(La)} &
\colhead{} & \colhead{[La/Fe]\tablenotemark{b}}
\\
\colhead{} & \colhead{} & \colhead{} & \colhead{(K)} &
\colhead{(${\rm cm\, s^{-2}}$)} & \colhead{} & \colhead{(${\rm
km\,s^{-1}}$)} & \colhead{} & \colhead{} & \colhead{} & \colhead{}
& \colhead{} & \colhead{} & \colhead{}& \colhead{}& \colhead{} &
\colhead{} & \colhead{}
 }

\startdata
Sun & \nodata & \nodata & \nodata & \nodata && \nodata & 7.45 & \nodata  & 4.90 & \nodata & &  2.21 &  \nodata & & 1.13 & & \nodata   \\
\\
TriAnd &  &  &&  &  &  &  &  &  &  &  &  &  & &  & \\
$J01214158+3635505$ & 10.65 & 1.06 & 3750 & 0.4&    & 1.71 & 6.86 & $-0.59/  0.11$ & 3.83 & $-0.48 / 0.13$    && 1.47 & $-0.15 / 0.05$   & &  0.30& & $-0.24 / 0.09$ \\
$J01425641+3851201$ & 10.73 & 1.07 & 3750 & 0.3&    & 1.74 & 6.74 & $-0.71/  0.09$ & 4.33 & $ 0.14 / 0.05$    && 1.36 & $-0.14 / $\nodata&\tablenotemark{c} &  0.32& & $-0.10 / 0.02$ \\
$J02044137+4059528$ & 11.11 & 1.00 & 3850 & 0.4&    & 1.74 & 6.54 & $-0.91/  0.11$ & 4.02 & $ 0.03 / 0.02$    && 1.27 & $-0.03 / $\nodata&\tablenotemark{c} &  0.21&\tablenotemark{d} & $-0.01 / 0.02$ \\
$J23333834+3909235$ & 10.60 & 1.07 & 3750 & 0.4&    & 1.66 & 6.82 & $-0.63/  0.11$ & 3.94 & $-0.33 / 0.01$    && 1.33 & $-0.25 / 0.21$    &&  \nodata&\tablenotemark{e} & \nodata$/$\nodata  \\
$J23490540+4057312$ & 11.11 & 1.02 & 3850 & 0.9&    & 1.53 & 7.12 & $-0.33/  0.14$ & 4.35 & $-0.22 / 0.20$    && 1.69 & $-0.19 / 0.12$    &&  0.82& & $0.02 / 0.09$ \\
$J23534927+3659173$ & 11.02 & 1.07 & 3750 & 0.4&    & 1.65 & 6.81 & $-0.64/  0.11$ & 3.85 & $-0.41 / 0.13$    && 1.61 & $ 0.04 / 0.12$    &&  0.15& & $-0.34 / 0.03$ \\

\enddata

\tablenotetext{a}{~The effective temperature as derived from the
Houdashelt et al. (2000) color-temperature
relation.}\tablenotetext{b}{~Abundances are shown with the
standard deviation in measurements. }\tablenotetext{c}{~Only one
\ion{Y}{+2} line measurable in two adjacent orders. }
\tablenotetext{d}{~Measurement uncertain due to spectrum defect on
the blue edge of the observed La line.} \tablenotetext{e}{~Lines
unmeasurable due to cosmic rays or other defects.}

\end{deluxetable}

\begin{figure}
\plotone{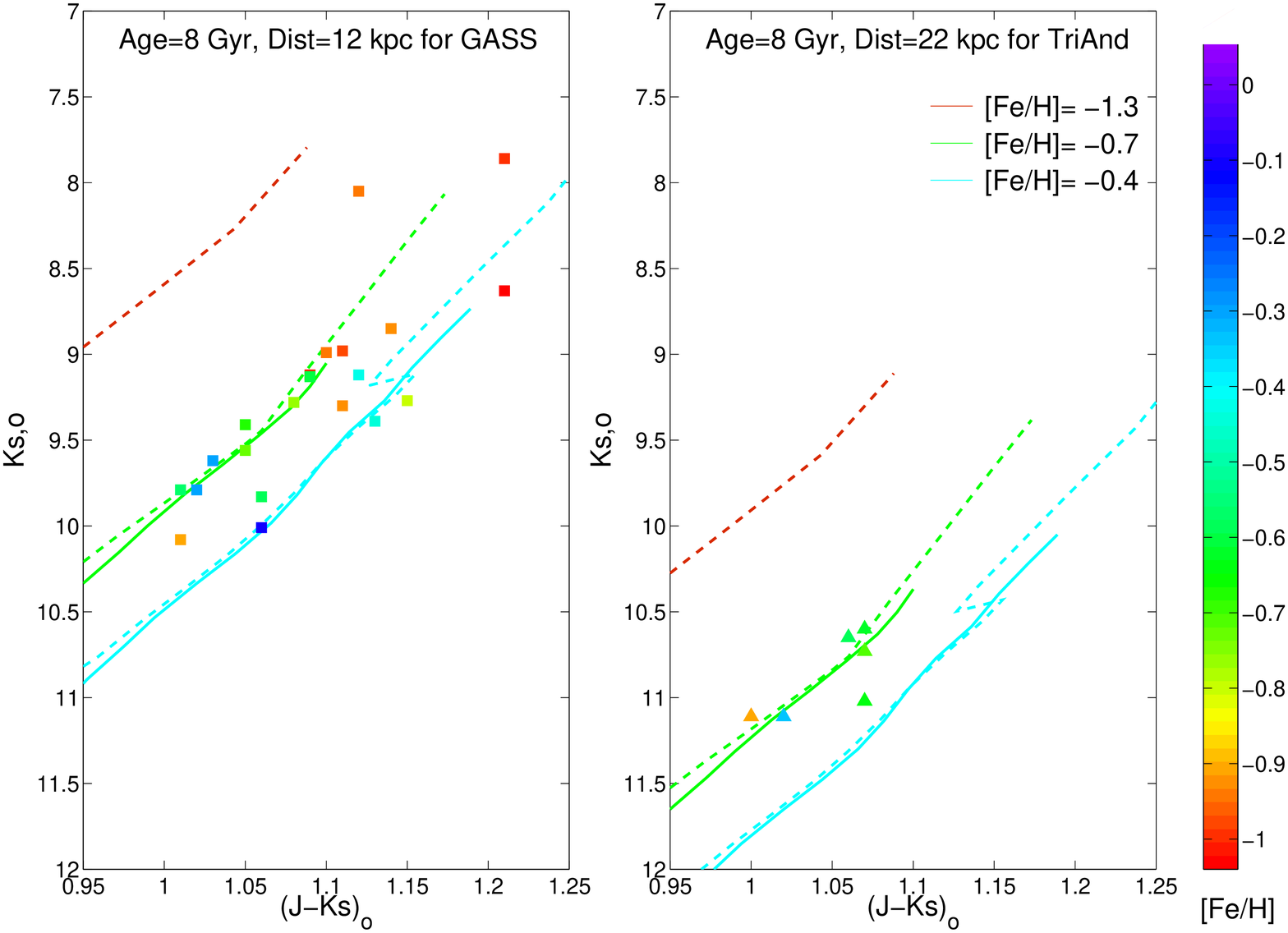} \caption{CMD of GASS (left) and TriAnd (right)
candidates compared with [Fe/H]$=-0.4$, $-0.7$ and $-1.3$
isochrones (Marigo et al. 2008) of age $=8$ Gyr, [$\alpha$/Fe] =
0.0, and distances $=12$ and 22 kpc, respectively. Solid lines are
RGB tracks and dashed lines are AGB tracks for the isochrones. The
color scale indicates derived [Fe/H] values for TriAnd and GASS
stars.}
\end{figure}

\begin{figure}
\plotone{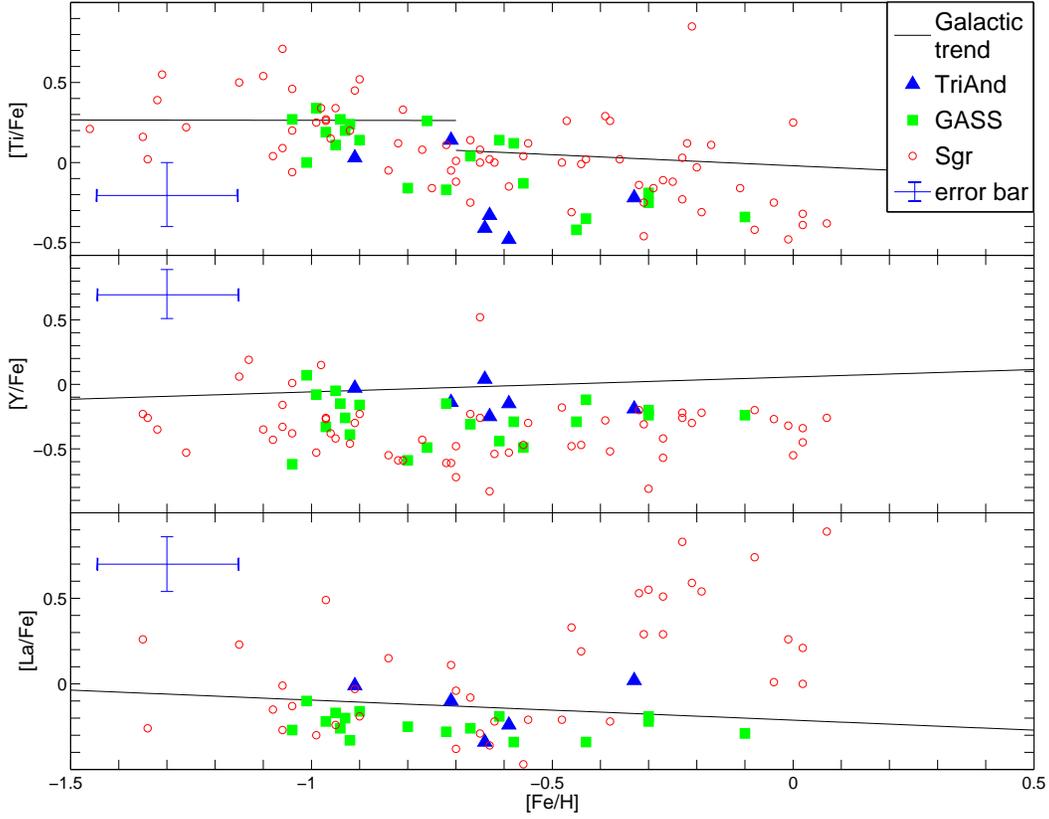} \caption{Distribution as a function of [Fe/H] of
(a) [Ti/Fe], (b) [Y/Fe] and (c) [La/Fe] for TriAnd (blue
triangles), GASS (green squares) and Sgr (red open circles) stars.
Lines represent linear fits to the MW star distribution (see
discussion in C10a). Typical error bars are shown in each panel in
blue. While roughly 70\% of the Ti abundancs and 50\% of the La
abundances shown for Sgr stars in the abundance range $-0.5 <$
[Fe/H] $<0$ come from other sources (Monaco et al.\ 2005 and
Sbordone et al.\ 2007; see C10a), their trends mimic those found
at these metallicities for our own Sgr sample (C10a); thus, for
clarity, we do not distinguish these various Sgr subsamples.}
\end{figure}

\end{document}